\documentclass[12pt]{article}

\usepackage{amsmath}

\def\eq#1{Eq.~(\ref{#1})}
\newcommand{\secn}[1]{Section~\ref{#1}}

\newcommand{\ord}{{\cal O}}
\newcommand{\RE}{{\rm Re}}

\def\beq{\begin{equation}}
\def\eeq{\end{equation}}
\def\beqa{\begin{eqnarray}}
\def\eeqa{\end{eqnarray}}
\def\ifm{\ifmmode}

\def \as {\relax\ifmmode\alpha_s\else{$\alpha_s${ }}\fi}

\def \al #1 {\frac {\as({#1})}{\pi} }
\def \ds #1 {\ooalign{$\hfil/\hfil$\crcr$#1$}}

\def\eps{\varepsilon}
\def\eq#1{Eq.~(\ref{#1})}

\begin{document}

\begin{titlepage}

\rightline{LPN13-086}

\vspace{1cm}

\centerline{\Large \bf High-energy QCD amplitudes}
\vspace{2mm}
\centerline{\Large \bf at two loops and beyond} 

\vspace{1.5cm}

\centerline{\bf Vittorio Del Duca\footnote{e-mail: {\tt Vittorio.Del.Duca@cern.ch}}}
\centerline{\sl Istituto Nazionale di Fisica Nucleare, Laboratori Nazionali di Frascati,}
\centerline{\sl Via E. Fermi 40, I-00044 Frascati, Italy}

\vspace{5mm}

\centerline{\bf Giulio Falcioni\footnote{e-mail: {\tt falcioni@to.infn.it}}, 
Lorenzo Magnea\footnote{e-mail: {\tt lorenzo.magnea@unito.it}} and 
Leonardo Vernazza\footnote{e-mail: {\tt vernazza@to.infn.it}}}
\centerline{\sl Dipartimento di Fisica, Universit\`a di Torino}
\centerline{\sl and INFN, Sezione di Torino}
\centerline{\sl Via P. Giuria 1, I--10125 Torino, Italy}

\vspace{1.5cm}
 
\begin{abstract}

\noindent Recent progress in our understanding of infrared singularities of multi-parton 
amplitudes has shown that the simplest form of Regge factorization for high-energy
gauge-theory amplitudes fails starting at next-to-next-to-leading logarithmic accuracy. 
We provide a framework to organize the calculation of parton amplitudes at leading
power in $t/s$, in terms of factorizing and non-factorizing contributions. This allows us 
to give explicit expressions for the leading Reggeization-breaking terms in two-loop
and three-loop quark and gluon amplitudes in QCD. In particular, using only infrared
information, we recover a known non-factorizing, non-logarithmic double-pole 
contribution at two-loops, and we compute the leading non-factorizing single-logarithmic
contributions at three loops.

\end{abstract}

\end{titlepage}

\newpage

\section{Introduction}
\label{intro}

In the high-energy limit, in which the centre-of-mass energy $\sqrt{s}$ is much larger 
than the typical momentum transfer $\sqrt{-t}$, so that $|s/t|\to \infty$, with $t$ held 
fixed, gauge theory scattering amplitudes become very simple: they acquire a factorized
structure, where the building blocks are given by a $t$-channel propagator, connecting
two emission vertices, often called {\it impact factors}, characterizing the particles 
undergoing the scattering. This structure is often referred to as {\it high-energy 
factorization}: impact factors depend on the specific scattering process, but they have
a simple coupling to the $t$-channel propagator, which is process independent.

Going from tree level to loop corrections, the picture remains the same, but the 
$t$-channel propagator gets dressed according to the schematic form~\cite{Balitsky:1979ap},
\beq
  \frac1{t} \to \frac1{t} \left( \frac{s}{-t}\right)^{\alpha(t)}\,,
\label{eq:regge}
\eeq
where $\alpha(t)$ is a function of the coupling constant, which in the weak coupling 
limit becomes a series expansion in the coupling. Because of the analytic structure
of \eq{eq:regge}, which is typical of Regge theory, $\alpha(t)$ is called {\it Regge
trajectory}.

Since the amplitude has a $t$-channel ladder-like structure, we can assume it
to be even under $s \leftrightarrow u$ exchange. As a consequence, it must be 
composed of kinematic and color parts which are either both even or both odd 
under $s \leftrightarrow u$ exchange. If one considers $t$-channel gluon exchange, 
which is all that is needed at leading order and at leading logarithmic accuracy in 
$\ln(s/|t|)$, then one takes the amplitude to be composed of kinematic and color 
parts which are both odd under $s\leftrightarrow u$ exchange. To be definite, let 
us consider the amplitude for gluon-gluon scattering. In this case, for the process
$g(k_1) + g(k_2) \to g(k_3) + g(k_4)$, one may write~\cite{Fadin:1993wh}
\beqa
  && {\cal M}^{g g \rightarrow g g}_{a_1 a_2 a_3 a_4} \left(\frac{s}{\mu^2}, 
  \frac{t}{\mu^2}, \alpha_s (\mu^2) \right)
  \, = \, 4 \pi \alpha_s (\mu^2)  \, \, \frac{s}{t} \, \bigg[ (T^b)_{a_1 a_3}
  C_{\lambda_1\lambda_3}(k_1, k_3) \bigg] \nonumber \\ 
  && \hspace{1cm} \times \,\, \left[ \left( 
  \frac{s}{- t} \right)^{\alpha(t)} \!\! + \left( \frac{- s}{- t} 
  \right)^{\alpha(t)} \right]
  \bigg[ (T^b)_{a_2 a_4} C_{\lambda_2\lambda_4}
  (k_2, k_4) \bigg] \, .
\label{Mgg2}
\eeqa
where $a_j$ and $k_j$ are the color index and momentum of gluon $j$, and 
$T^b$ is a color generator in the adjoint representation, so that $(T^a)_{b c} = 
- {\rm i} f_{a b c}$. The impact factors, $C_{\lambda_i\lambda_j}(k_i, k_j)$, 
depend on the  helicities of the gluons, but, as the notation suggests, carry 
no $s$~dependence. Both the impact factors and the Regge trajectory, in 
the weak coupling limit, can be expanded in powers of the renormalized
coupling $\alpha_s(\mu^2)$: they are then affected by infrared and collinear 
divergences, which in \eq{Mgg2} are (implicitly) regularized by dimensional 
regularization.

Beyond leading order, one should consider also the exchange of two or more 
reggeized gluons. Accordingly, one must include the contribution to the amplitude 
in which the kinematic and color parts are both even under $s \leftrightarrow u$ 
exchange, and in particular the case in which a color singlet is exchanged. 
\eq{Mgg2}, however, suffices to describe the amplitude at leading and at 
next-to-leading logarithmic (NLL) accuracy in $\ln(s/|t|)$~\cite{Fadin:2006bj}.

By writing formulae similar to \eq{Mgg2} for quark-quark and quark-gluon scattering,
and considering them together with gluon-gluon scattering as given by \eq{Mgg2}, 
one obtains a system of three equations. Their expansion at one loop shows that 
each equation has a term proportional to $\ln(s/|t|)$, which is the same for all three 
amplitudes. That term gives the one-loop Regge trajectory, and the fact that is the same 
for all three equations shows its universality, {\it i.e.} its independence of the particular 
scattering process under consideration. Conversely, the term independent of $\ln(s/|t|)$
is different for each equation. Thus one gets an over-constrained system of three 
coefficients and two unknowns, the one-loop impact factors for quark and gluon 
scattering. One can use two of the coefficients to determine the one-loop impact 
factors, and the third to perform a consistency check on high-energy factorization. 
Repeating the same procedure at two loops, one can use the terms proportional to 
$\ln(s/|t|)$ to determine the two-loop Regge trajectory and verify its universality, and 
the terms independent of $\ln(s/|t|)$ to compute the two-loop impact factors and check 
that high-energy factorization holds. Such a check, however,  fails~\cite{DelDuca:2001gu}, 
due to the presence of a term proportional to $\alpha_s^2 \pi^2/\epsilon^2$, which 
therefore invalidates high-energy factorization, making the determination of the 
two-loop impact factors ambiguous.

A general approach to the high-energy limit of gauge theory amplitudes based on the
universal properties of their infrared singularities, developed in~\cite{DelDuca:2011ae,
Bret:2011xm}, following the earlier results of~\cite{Korchemsky:1993hr,Korchemskaya:1994qp,
Korchemskaya:1996je}, suggests that the violation of high-energy factorization 
reported in~\cite{DelDuca:2001gu} at order $\alpha_s^2$ and at next-to-next-to-leading 
logarithmic accuracy in $\ln(s/|t|)$ is due to the amplitude becoming non-diagonal in the 
$t$-channel-exchange-basis. Such a violation iterates then at three loops in the $\alpha_s^3$ 
term proportional to $\ln(s/|t|)$, invalidating the universality of the three-loop Regge trajectory.
Thus, the eventual definition of a universal three-loop Regge trajectory requires
additional conditions.

The goal of this letter is to pinpoint the origin of the high-energy factorization violation 
discovered in~\cite{DelDuca:2001gu} at two loops, and to propose a way to isolate
factorization-breaking terms at three loops and beyond, in order to be able to define
unambiguously a universal Regge trajectory and the related impact factors. This implies 
the definition of a non-factorizing contribution to the amplitude, whose infrared and 
collinear divergent part can then be unambiguously predicted using the tools described
in~\cite{DelDuca:2011ae,Bret:2011xm}. We believe that a framework for consistently 
identifying factorizing and non-factorizing contributions to high-energy amplitudes can 
be useful both in practical finite-order calculations, to assess the reliability of high-energy 
resummations, and for theoretical developments. Indeed, a violation of na\"ive 
high-energy factorization, as given for example by~\eq{Mgg2}, at NNLL accuracy and 
for non-planar contributions to the amplitude, could have been predicted in the context of 
Regge theory~\cite{Collins:1977jy} by noting that at this level one may expect contributions 
to the amplitude due to Regge cuts in the angular momentum plane, whereas expressions 
of the form of \eq{Mgg2} arise under the assumption that the only singularities in the 
$l$ plane be isolated poles. A precise expression for the discrepancy between pole-based
Regge factorization and the actual perturbative results for the amplitude may be useful
at least as a boundary condition for future attempts to extend high-energy factorization
to include the contributions of Regge cuts. Furthermore, our results are a first step
in the direction of systematically combining informations on high-order amplitudes
which arise from soft-collinear factorization, which is exact to all orders in perturbation 
theory for all singular contributions to the amplitudes, with those arising from Regge
factorization, which apply to finite contributions to the amplitudes as well, but have 
limited validity in terms of logarithmic accuracy. The combination of the two approaches,
within the framework discussed in the present letter, yields towers of constraints on 
real and imaginary parts of finite order amplitudes, which we will discuss in detail in a 
forthcoming publication~\cite{us}.

In the following, we begin by briefly reviewing, in \secn{irhe}, the results of 
Ref.~\cite{DelDuca:2011ae}, in order to set up our notation in a general context.
In \secn{hepa}, we provide a general parametrization of four-point quark and gluon
amplitudes in the high-energy limit, which we then use in \secn{coirf} to compare
in detail the two factorizations. This allows us to recover the results of~\cite{DelDuca:2001gu},
and to provide a definite prediction for factorization-breaking terms at three loops.
We conclude by briefly discussing the results and the prospects for future developments
in \secn{discu}.

\section{Infrared divergences at high-energy}
\label{irhe}

We consider a scattering process of $2 \to 2$ massless on-shell partons. Each 
parton carries a color index, and we may write the scattering amplitude as a 
vector in color space,
\beq
  {\cal M}_{2 \to 2}^{a a' b b'} \left(\frac{s}{\mu^2}, \frac{t}{\mu^2}, \as(\mu^2) \right) \, =  \,
  \sum_{j} {\cal M}_{2 \to 2}^{[j]} \left( \frac{s}{\mu^2}, \frac{t}{\mu^2}, \as(\mu^2) \right) \,
  c^{a a' b b'}_{[j]} \, ,
\label{ColExpansion}
\eeq
where the index $[j] = 1, \ldots, r$ runs over the color representations which are allowed 
in a given channel exchange, and $c^{aa'bb'}_{[j]}$ is a suitable orthonormal basis of 
color tensors. For a detailed discussion of how such tensors can be enumerated and 
constructed when the external particles are in generic color representations, we 
refer the reader to~\cite{DelDuca:2011ae}. As before, and as in the rest of the paper, in 
\eq{ColExpansion} we leave implicit the dependence on the infrared regulator $\epsilon 
= 2 - d/2 < 0$.

The structure of infrared and collinear singularities of multi-parton amplitudes  
can be described, at least to the accuracy required in the present paper, by means 
of the dipole formula~\cite{Becher:2009cu,Gardi:2009qi,Becher:2009qa,Gardi:2009zv}.
This result is based on the factorization theorem for soft singularities of fixed-angle 
multi-parton scattering amplitudes (see for example~\cite{Dixon:2008gr} and references 
therein), which in this case can be written as
\beq 
  {\cal M}_{2 \to 2} \left(\frac{s}{\mu^2}, \frac{t}{\mu^2}, \as \right) \, = \, 
  {\cal Z}_{2 \to2} \left(\frac{s}{\mu^2}, \frac{t}{\mu^2}, \as \right)
  {\cal H}_{2 \to2} \left(\frac{s}{\mu^2}, \frac{t}{\mu^2}, \as \right) \, .
\label{IRfact}
\eeq
Here ${\cal H}$ is a color vector, finite as $\epsilon \rightarrow 0$, and representing a
matching condition to be determined order by order in perturbation theory after
subtraction of all infrared divergent contributions. It can be expressed in the same color 
basis as the full amplitude, as
\beq
  {\cal H}^{a a' b b'}_{2 \to 2} \left(\frac{s}{\mu^2}, \frac{t}{\mu^2}, \as \right) \, = \, 
  \sum_{j} {\cal H}_{2 \to 2}^{[j]} \left(\frac{s}{\mu^2}, \frac{t}{\mu^2}, \as \right)
  \, c^{a a' b b'}_{[j]} \, .
\label{HColExpansion}
\eeq
On the other hand, ${\cal Z}$ is an $r \times r$ matrix in color space, with matrix 
elements ${\cal Z}_{[j],[j']}$ and $j,j'$ running over the $r$ allowed color representations
in the selected channel. ${\cal Z}$ generates all the infrared and collinear singularities 
of the amplitude. As detailed in Refs.~\cite{Gardi:2009qi,Becher:2009qa}, it can be written 
in full generality, for $2 \to n$ parton scattering, in terms of an anomalous dimension 
matrix $\Gamma$ as
\beq
  {\cal Z}_{2 \to n} \left(\frac{p_i}{\mu}, \as \right) \, = \,  
  {\cal P} \exp \left[ \frac{1}{2} \int_0^{\mu^2} \frac{d \lambda^2}{\lambda^2} \, \,
  \Gamma_{2 \to n} \left(\frac{p_i}{\lambda}, \alpha_s(\lambda^2) \right) \right] \, ,
\label{RGsol}
\eeq
where ${\cal P}$ denotes path ordering in color space, and all singularities in $\epsilon$
are generated through the integration of the $d$-dimensional running coupling down
to vanishing scale $\lambda \to 0$. The results of Refs.~\cite{Aybat:2006mz,Gardi:2009qi,
Becher:2009qa} show that, at least up to two loops, the anomalous dimension matrix
takes the form
\beq
  \Gamma_{2 \to n}^{\rm dip}  \left(\frac{p_i}{\lambda}, \alpha_s(\lambda^2) \right) \, = \,
  \frac{1}{4} \, \widehat{\gamma}_K \left(\alpha_s (\lambda^2) \right) \,
  \sum_{(i,j)} \ln \left(\frac{- s_{i j}}{\lambda^2} 
  \right) {\bf T}_i \cdot {\bf T}_j \, - \, \sum_{i = 1}^{n + 2}
  \gamma_{J_i} \left(\alpha_s (\lambda^2) \right) \, .
\label{sumodipoles}
\eeq
The basic feature of \eq{sumodipoles} is that the color structure, expressed in terms
of color-insertion operators ${\bf T}^i$ appropriate to the color representation of hard
parton $i$, is simply expressed as a sum over color dipoles, with all higher-order multipoles
vanishing exactly. Color degrees of freedom are tightly correlated with kinematics,
through the invariants $s_{i j} = (p_i + p_j)^2$, where for the sake of simplicity we have 
taken all momenta as outgoing. Since the color structure in \eq{sumodipoles} is fixed
at one loop, the path ordering symbol in \eq{RGsol} can be dropped when employing 
\eq{sumodipoles}. All dependence on the coupling is confined to colorless anomalous
dimensions: $\hat{\gamma}_K = \gamma_K^{[i]}/{\cal C}_{[i]}$, where $\gamma_K^{[i]}$
is the cusp anomalous dimension~\cite{Korchemsky:1985xj,Korchemsky:1987wg} 
in representation $[i]$ and ${\cal C}_{[i]}$ is the corresponding quadratic Casimir 
eigenvalue, and the collinear anomalous dimensions $\gamma_{J_i}$, which can be
extracted from form factor data~\cite{Becher:2009qa,Dixon:2008gr,Magnea:1990zb}.

The dipole formula, \eq{sumodipoles}, is exact up to two loops for massless partons.
Possible corrections beyond two loops have been studied in detail in~\cite{Becher:2009qa,
Dixon:2009ur,Ahrens:2012qz}: they can only take the form of tightly constrained conformal
cross-ratios of kinematic invariants, starting at three loops and with at least four hard partons,
or they can arise as a consequence of violations of Casimir scaling for the cusp anomalous
dimension, which can happen in principle starting at four loops. The exact calculation of the
three-loop soft anomalous dimension matrix $\Gamma$ is a vastly challenging project, and
recent progress to this end has very recently been summarized in~\cite{Gardi:2013saa}. 
Also very recently, evidence for a failure of the dipole formula at the four-loop level, and 
at NLL accuracy in the high-energy limit, was provided in~\cite{Caron-Huot:2013fea}. While 
these are very interesting results, obtained with innovative techniques, they do not influence 
the outcome of our calculations, which only concern terms that are fully accounted for by 
the dipole ansatz\footnote{Notice in particular that the violation of the dipole ansatz discussed
in~\cite{Caron-Huot:2013fea} arises in the terms of the amplitude which are even both in color
and in kinematics, and which arise only when at least two Reggeized gluons are exchanged, 
whereas we focus on the terms the are odd in both sets of variables.}.

In the high energy limit, $s/|t| \to \infty$, the four-point scattering amplitude is affected by
large logarithms $\ln(s/(-t))$, which are the focus of our investigation. To leading power
in $t/s$, the amplitude can then be organized as a double expansion, in the coupling constant
and in the power of the large logarithm. For each color component of the vector ${\cal M}$ 
we write
\beq
  {\cal M}_{2 \to 2}^{[j]} \left(\frac{s}{\mu^2}, \frac{t}{\mu^2}, \as \right) \, = \, 4 \pi \as \, 
  \sum_{n = 0}^\infty \sum_{i = 0}^n
  \left( \frac{\as}{\pi} \right)^n \ln^i \left( \frac{s}{- t} \right)
  M^{(n), i, [j]} \left( \frac{t}{\mu^2} \right) \, ,
\label{AmpExpansion}
\eeq
with corrections suppressed by powers of $t/s$. The components of the finite hard
vector ${\cal H}$ can be expanded likewise,
\beq
  {\cal H}_{2 \to 2}^{[j]} \left( \frac{s}{\mu^2}, \frac{t}{\mu^2}, \as \right) \, = \, 
  4 \pi \as \,\sum_{n = 0}^\infty \sum_{i = 0}^n
  \left( \frac{\as}{\pi} \right)^n \ln^i \left( \frac{s}{-t} \right)
  H^{(n), i, [j]} \left( \frac{t}{\mu^2} \right) \, .
\label{HExpansion}
\eeq
The matrix ${\cal Z}$, on the other hand, was shown in~\cite{DelDuca:2011ae,Bret:2011xm}
to factorize, to leading power in $t/s$, according to
\beq
  {\cal Z} \left(\frac{s}{\mu^2}, \frac{t}{\mu^2}, \as \right) \, = \, 
  {\cal Z}_1 \left(\frac{t}{\mu^2}, \as \right)
  \widetilde{\cal Z} \left(\frac{s}{t}, \as \right) + \ord \left(\frac{t}{s} \right) \, ,
\label{Zfact}
\eeq
where
\beq
  \widetilde{{\cal Z}} \left(\frac{s}{t}, \as \right)
  \, = \, \exp \left\{ K( \as )
  \left[ \log \left( \frac{s}{-t } \right) {\bf T}_t^2 + {\rm i} \pi {\bf T}_s^2\right] \right\}
\label{widetildeZ}
\eeq
is a matrix in the same color space as ${\cal Z}$, and is responsible for generating all
the large logarithms of the amplitude which are accompanied by infrared poles. In 
\eq{widetildeZ} we have introduced the `Mandelstam' combinations of color-insertion
operators ${\bf T}_s = {\bf T}_1 + {\bf T}_2$ and ${\bf T}_t = {\bf T}_1 + {\bf T}_3$. The 
coefficients of the high-energy logarithms are determined by the function
\beq
  K \left( \as \right) = 
  - \frac{1}{4} \int_0^{\mu^2} \frac{d \lambda^2}{\lambda^2} \,
  \hat{\gamma}_K \left( \alpha_s (\lambda^2) \right) \,,
\label{cusp}
\eeq
which is a scale integral over the cusp anomalous dimension\footnote{This integral plays an
important and ubiquitous role in perturbative QCD: the dimensionally regularized version
in \eq{cusp} emerged first in the resummation of infrared poles in the quark form factor
in~\cite{Magnea:1990zb} and was recursively computed to all orders, in terms of the 
perturbative coefficients of $\beta(\as)$ and $\gamma_K (\as)$, in~\cite{Magnea:2000ss}. 
In the context of the high-energy limit a slightly different form of \eq{cusp} was shown to 
give the all-order infrared part of the Regge trajectory in~\cite{Korchemskaya:1996je}.}. 
In \eq{cusp} the (singular) $\epsilon$ dependence is generated through integration of 
the $d$-dimensional version of the running coupling, so that the result is a pure counterterm, 
easily computed order by order in terms of the perturbative coefficients of the the $\beta$ 
function and of the cusp anomalous dimension. To two-loop order one finds for example
\beqa
  K (\as) & = & \frac{\alpha_s}{\pi} \, 
  \frac{\widehat{\gamma}_K^{(1)}}{4 \epsilon} \, + \left(\frac{\alpha_s}{\pi}\right)^2 \,
  \left( \frac{\widehat{\gamma}_K^{(2)}}{8 \epsilon} -
  \frac{b_0 \, \widehat{\gamma}_K^{(1)}}{32 \epsilon^2} \right) + \ord (\alpha_s^3) \, .
\label{KNNLO}
\eeqa
Note that the elements of the matrices ${\cal Z}$ and $\widetilde{{\cal Z}} $ in Eqs.~(\ref{Zfact}) 
and (\ref{widetildeZ}) may be written as double expansions in the coupling constant
and in the large logarithms, as was done in Eqs.~(\ref{AmpExpansion}) and (\ref{HExpansion}).

As shown explicitly in Ref.~\cite{DelDuca:2011ae}, \eq{widetildeZ} can be used as a starting 
point to analyze the all-order structure of high-energy logarithms accompanied by infrared
poles. To leading logarithmic (LL) accuracy, one easily recovers the Reggeization of the
parton exchanged in the $t$ channel. \eq{widetildeZ} is however valid to all logarithmic 
orders at leading power, and one can use it to study Reggeization and its breaking beyond 
LL. For example, one finds that at NNLL non-Reggeizing logarithms must appear starting
at three loops, with the leading effects arising from the operator
\beq
  {\cal E} \left( \frac{s}{t}, \alpha_s \right) \equiv - \, \frac{\pi^2}{3} \,
  {K^3 (\alpha_s )} \, \ln \left(\frac{s}{- t} \right) 
  \Big[{\bf T}_s^2, \big[{\bf T}_t^2, {\bf T}_s^2 \big] \Big] \, ,
\label{NNLL}
\eeq
One of the goals of this letter is to evaluate explicitly the effect of this operator at three 
loops in quark and gluon amplitudes.

Turning back to \eq{Zfact}, the remaining factor ${\cal Z}_{\bf 1}$ is a singlet in color 
space, and we write it explicitly here as
\beq
  {\cal Z}_1 \left(\frac{t}{\mu^2}, \as \right) \, = \,
  {\cal Z}_{1, {\bf R}} \left(\frac{t}{\mu^2}, \as \right) \, 
  \exp \left( - {\rm i} \, \frac{\pi}{2} \, K \left( \as \right)
  {\cal C}_{\rm tot} \right) \, ,
\label{facZ1}
\eeq
where we have isolated the phase factor, expressed in terms of the cusp and of the
combined Casimir eigenvalue ${\cal C}_{\rm tot} \equiv \sum_{i = 1}^{4} {\cal C}_{i}$,
leaving behind a function which is real in the physical region, and which in turn is given 
by
\beqa
  {\cal Z}_{1, {\bf R}} \left(\frac{t}{\mu^2}, \as \right) \! & = & \!
  \exp \Bigg\{  \frac{1}{2} \Bigg[ K \left( \as \right) \log \left( \frac{-t}{\mu^2} 
  \right) + D \left( \as \right) \Bigg] {\cal C}_{\rm tot}  
  + \, \sum_{i = 1}^4 B_i \left( \as \right) \Bigg\} \,, \nonumber\\
  & = & \sum_{n = 0}^{\infty} \left( \frac{\as}{\pi} \right)^n
  Z_{1, {\bf R}}^{(n)} \left( \frac{t}{\mu^2} \right) \, ,
\label{Z1}
\eeqa
where in the second line we have written ${\cal Z}_{1, {\bf R}}$ as an expansion over
the coupling constant. The functions $D(\as)$ and $B(\as)$ are given by scale 
integrals over the cusp and collinear anomalous dimensions, as in \eq{cusp}, and they 
similarly yield a perturbative series of pure counterterms, representing infrared and collinear
divergences. Explicitly,
\beqa
  D \left( \as \right) & = &
  - \frac{1}{4} \int_0^{\mu^2} \frac{d \lambda^2}{\lambda^2} \,
  \widehat{\gamma}_K \left( \alpha_s (\lambda^2) \right) 
  \log \left(\frac{\mu^2}{\lambda^2}\right), \nonumber\\
  B_i \left( \as \right) & = &
  - \frac{1}{2} \int_0^{\mu^2} \frac{d \lambda^2}{\lambda^2} \,
  \gamma_{J_i} \left(\alpha_s (\lambda^2) \right) \,.
\label{intandim}
\eeqa
An important property of the operator ${\cal Z}_{{\bf 1}, {\bf R}}$, and indeed of 
${\cal Z}_{{\bf 1}}$, is that, to all orders, it is the product of four factors, each one 
associated with one of the external hard partons. One may write
\beq
  {\cal Z}_{{\bf 1}, {\bf R}} \left( \frac{t}{\mu^2}, \as \right) 
  \, = \, \prod_{i = 1}^4 {\cal Z}_{{\bf 1},{\bf R}}^{(i)} \left( \frac{t}{\mu^2}, \as \right)\, ,
\label{jetfactors}
\eeq
and similarly for ${\cal Z}_{{\bf 1}}$. Each factor ${\cal Z}_{{\bf 1},{\bf R}}^{(i)}$ is thus
properly thought of as a `jet' operator, and one may expect these jet operators to
combine naturally to yield the divergent parts of the impact factors. We will see
below that this is indeed the case.

\section{The structure of high-energy parton amplitudes}
\label{hepa}

In \eq{Mgg2}, we have displayed the Regge factorization formula for gluon-gluon 
scattering, with the $t$-channel exchange of a reggeized gluon. In order to include also 
quark-quark and quark-gluon scattering, we need to take into account the fact that
the color factor for the quark-quark amplitude does not have a definite symmetry 
property under $s \leftrightarrow u$. In that case, therefore, the symmetric and the 
antisymmetric parts of the kinematic factor must have different weights. We write then,
for the octet component of the matrix element,
\beqa
  {\cal M}_{ab}^{[8]} \left(\frac{s}{\mu^2}, \frac{t}{\mu^2}, \as \right)
  & = & 2 \pi \alpha_s \, H^{(0),[8]}_{ab}  
  \nonumber \\ && \hspace{-2cm} \times \,\,
  \Bigg\{
  C_a \left(\frac{t}{\mu^2}, \as \right)
  \bigg[ A_+ \left(\frac{s}{t}, \as \right) + \, \kappa_{ab} \,
  A_- \left(\frac{s}{t}, \as \right) \bigg]
  C_b \left(\frac{t}{\mu^2}, \as \right) \nonumber \\
  & & \, + \, \, {\cal R}_{ab}^{[8]} \left(\frac{s}{\mu^2}, \frac{t}{\mu^2}, \as \right)
 + \ord \left( \frac{t}{s} \right) \Bigg\} \, ,
\label{ReggeFact}
\eeqa
where the indices $a,b$ label the parton species (quark or gluon), and
\beq
  A_\pm \left(\frac{s}{t}, \as \right) \, = \, \left( \frac{- s}{- t} \right)^{\alpha(t)}
  \pm \left( \frac{s}{-t} \right)^{\alpha(t)} \, , 
\label{ReggeStructure}
\eeq
while $\kappa_{gg} = \kappa_{qg} = 0$, and $\kappa_{qq} = (4 - N_c^2)/N_c^2$.
In \eq{ReggeFact}, we can expand the Regge trajectory and the impact factors in 
powers of the coupling constant, as
\beq
  \alpha(t) \, = \, \sum_{n = 1}^{\infty} \left( \frac{\as}{\pi} \right)^n \alpha^{(n)}(t) \, ,
  \qquad C_i \left(\frac{t}{\mu^2}, \as \right) \, = \, \sum_{n = 0}^{\infty} 
  \left( \frac{\as}{\pi} \right)^n C_i^{(n)} \left(\frac{t}{\mu^2} \right)  \, ,
\label{expalc}
\eeq
and we have chosen the prefactor (where $H^{(0),[8]} = H^{(0),0,[8]}$ in the notations 
of \eq{HExpansion}) so that $C_i^{(0)} = 1$. If one had included only the first line in 
braces in \eq{ReggeFact}, the resulting expression would have been accurate only to 
NLL, and only for the real part of the amplitude. In order to promote the equality to leading 
power accuracy, we have included a non-factorizing remainder, ${\cal R}_{ab}^{[8]}$,
collecting all terms in the matrix element which cannot be written in terms of a universal
Regge trajectory with impact factors depending only on the parton species. We know
from earlier results that the non-factorizing remainder starts at two loops and at NNLL 
level, therefore we expand it in powers of the coupling and of the high-energy logarithm as
\beq
  {\cal R}_{ab}^{[8]} \left(\frac{s}{\mu^2}, \frac{t}{\mu^2}, \as \right)
  \, = \, \sum_{n = 2}^{\infty} \sum_{k = 0}^{n - 2}
  \left( \frac{\as}{\pi} \right)^n \ln^k \left( \frac{s}{- t} \right)
  R_{\, ab}^{\, (n), k, [8]} \left(\frac{t}{\mu^2} \right) \,.
\label{Rexp}
\eeq
Clearly, as with any factorization which breaks down at some level of accuracy, there
is a degree of ambiguity in the definition of the non-factorizing remainder ${\cal R}_{ab}^{[8]}$,
as it may be possible to move some (non-logarithmic) terms from the remainder to the impact 
factors without invalidating \eq{ReggeFact}. As we will see however, at least as far as infrared 
divergent contributions are concerned, the knowledge of the structure of the amplitude which 
comes from soft-collinear factorization provides us a very natural choice of `factorization 
scheme', and therefore with a natural choice for the non-factorizing remainder.

\section{Comparing soft-collinear and high-energy factorizations}
\label{coirf}

We now have at our disposal two different factorizations: \eq{IRfact}, with all the
subsidiary information collected in \secn{irhe}, and \eq{ReggeFact}. Soft-collinear 
factorization, embodied by \eq{IRfact}, is exact to all orders in perturbation theory
for infrared divergent contributions, and the high-energy limit of the ${\cal Z}$ matrix
is accurate to leading power in $t/s$. High-energy factorization as given in \eq{ReggeFact}
applies also to finite contributions to the amplitude, but has a limited logarithmic accuracy.
Our task is to intersect the informations from the two limits, extract the constraints 
that arise when both are applicable, and eventually make predictions based on one of them
when the second one breaks down.

To illustrate our strategy, we briefly summarize what happens at one loop, where all
ingredients are known and we are basically performing a consistency check. Throughout
this section we set $\mu^2 = - t$ so that all results for the trajectory and the impact factors
are given by pure numbers. We begin by expanding the available expressions for the matrix 
elements to first order in $\as$. For simplicity, we will omit the parton indices $a,b$ whenever 
they are not specifically needed. Soft-collinear factorization yields the expressions
\beqa
  M^{(1),0} & = & \left[ Z_{1, {\bf R}}^{(1)} + i \pi K^{(1)} \left({\bf T}_s^2  - 
  \frac{1}{2} {\cal C}_{\rm tot} \right) \right] H^{(0)} + H^{(1),0}, \nonumber \\
  M^{(1),1} & = & K^{(1)} {\bf T}_{t}^2 H^{(0)} + H^{(1),1} \, ,
\label{AmpCoeff1}
\eeqa
which are still vectors in color space, while for the octet component, high-energy
factorization provides the expressions
\beqa
  M^{(1),0,[8]}_{ab} & = & \left[ C_a^{(1)} + C_b^{(1)}
  - i \frac{\pi}{2} (1 + \kappa_{ab}) \alpha^{(1)} \right] H^{(0),[8]}_{a b} \, , \nonumber \\
  M^{(1),1, [8]}_{ab} & = & \alpha^{(1)} H^{(0), [8]}_{ab} \, .
\label{ReggeCoeff1}
\eeqa
One of the constraints of Regge factorization is the fact that the Regge trajectory
and the impact factors are required to be real: in other words, the imaginary part 
of the amplitude is completely determined by the `signature' properties under the
exchange $s \leftrightarrow u$, as given by \eq{ReggeFact} and by \eq{ReggeStructure}.
There are therefore interesting informations to be extracted about the imaginary parts
of the amplitude when comparing results such as \eq{AmpCoeff1} and \eq{ReggeCoeff1}.
Detailed results for imaginary parts will be discussed in~\cite{us}: here we focus on 
the real part of the amplitude. Comparing first one-loop terms proportional to $\ln(s/(-t))$,
we immediately see that we can write the one-loop Regge trajectory as
\beq
  \alpha^{(1)} \, = \, \frac{K^{(1)} \left( {\bf T}_t^2 H^{(0)} \right)^{[8]} }{H^{(0),[8]}} 
  + \frac{H^{(1),1,[8]}}{H^{(0),[8]}} \, .
\label{alpha1}
\eeq
In the high-energy limit, for all parton species, the tree-level amplitude is a pure color 
octet in the $t$-channel, and therefore it is an eigenvector of the ${\bf T}_t^2$ operator,
so that ${\bf T}_t^2 H^{(0)} = C_A H^{(0),[8]}$. Furthermore one easily verifies that 
$H^{(1),1,[8]} = {\cal O} (\epsilon)$. As expected, the Regge trajectory then becomes
\beq
  \alpha^{(1)} \, = \, C_A K^{(1)} + \ord(\eps) \, ,
\label{alpha1qqqggg}
\eeq
which confirms the universality of the one-loop Regge trajectory~\cite{Tyburski:1975mr,
Fadin:1975cb,Lipatov:1976zz,Kuraev:1976ge,Mason:1976fr,Cheng:1977gt,Kuraev:1977fs} 
to $\ord(\eps)$.

Turning to non-logarithmic contributions to the matrix elements in Eqs.~(\ref{AmpCoeff1}) 
and (\ref{ReggeCoeff1}), we can consider separately the quark-quark and the gluon-gluon 
scattering amplitudes, and determine the respective impact factors. One finds that
\beq
 C_a^{(1)} \, = \, \frac{1}{2} Z_{1, {\bf R}, a}^{(1)}  + \frac{1}{2} \widehat{H}^{(1),0,[8]}_{aa} \, ,
\label{imp1}
\eeq
where we defined  $\widehat{H}^{(m), n, [J]}_{ab} \, = \, H^{(m),n,[J]}_{ab}/H^{(0),[8]}_{ab}$, 
and where we have used the fact that, by virtue of \eq{jetfactors},
\beq
  Z_{1, {\bf R}, qg}^{(1)} \, = \,  \frac{1}{2} \bigg[ Z_{1, {\bf R}, qq}^{(1)} 
 + Z_{1, {\bf R}, gg}^{(1)} \bigg] \, .
\label{qgrelRE1}
\eeq
Having determined both  impact factors, one can finally verify the consistency of Regge 
factorization by constructing the high-energy quark-gluon scattering amplitude. One finds 
that requiring Regge factorization constrains the hard parts of the amplitudes to satisfy
\beq
  \RE \left( \widehat{H}^{(1), 0, [8]}_{qg} \right) \, = \, \frac{1}{2} \left[ \RE \left( 
  \widehat{H}^{(1), 0, [8]}_{gg} \right) + \RE \left( \widehat{H}^{(1), 0, [8]}_{qq} \right) \right] \, ,
\label{qgrelRE2}
\eeq
which is easily verified to be correct by using the explicit results listed, for example,
in Ref.~\cite{Kunszt:1993sd}.

Repeating the procedure at two loops, one finds more interesting results, and, at the level 
of non-logarithmic terms, one begins to see the breakdown of the high-energy factorization 
as given in \eq{ReggeFact}. Beginning at leading logarithms ($\ln^2 (s/(-t))$ at two loops), 
one readily verifies that the coefficient of the highest power of the energy logarithm is
determined by the one-loop result, as expected from high-energy resummation. At the 
level of single logarithms, comparing Eqs.~(\ref{ReggeFact}) and (\ref{AmpExpansion}) 
allows us to write the two-loop Regge trajectory~\cite{DelDuca:2001gu,Fadin:1995xg,Fadin:1996tb,
Fadin:1995km,Blumlein:1998ib} as
\beq
 \alpha^{(2)}  \,  =  \, C_A K^{(2)} + \RE \left[ \widehat{H}^{(2), 1, [8]}_{ab} \right]  + \ord(\eps) \, .
\label{alpha2}
\eeq
independently of the specific scattering process considered. This is again in perfect 
agreement with high-energy factorization. Turning to the terms which do not contain 
$\ln(s/(-t))$, however, we begin to see the effects of Reggeization breaking. In particular, 
deriving the two-loop quark and gluon impact factors from the factorized expression for 
the quark-quark and gluon-gluon scattering amplitudes respectively, we get, for the singular
terms of the impact factors,
\beqa
  C_a^{(2)} & = & \frac{1}{2} Z^{(2)}_{1, {\bf R}, aa} - \frac{1}{8} 
  \left( Z^{(1)}_{1, {\bf R}, aa} \right)^2 + \frac{1}{4} Z^{(1)}_{1, {\bf R}, aa} \, 
  \RE \left[ \widehat{H}^{(1), 0, [8]}_{aa} \right] - \frac{1}{4} R_{aa}^{(2), 0, [8]}
  \label{c2imp} \\ && \hspace{-17mm} 
  - \frac{\pi^2 (K^{(1)})^2}{4} \bigg\{ \left[ \left( {\bf T}^2_{s, aa} \right)^2 \right]_{[8], [8]}
  - {\cal C}_{{\rm tot}, aa} \left[ {\bf T}^2_{s, aa} \right]_{[8], [8]}
  + \frac{1}{4} {\cal C}^2_{{\rm tot}, aa} - \frac{(1 + \kappa_{aa}) C_A^2}{2} \bigg\} 
  \, , \nonumber 
\eeqa
with $a = q,g$, and where we have allowed for a non-vanishing non-factorizing
remainder $R$, according to \eq{ReggeFact}.

We observe that soft-collinear factorization has generated an expression for the
impact factors which manifestly contains both universal and non-universal components.
Indeed, the first line of \eq{c2imp}, with the exception of the so-far undefined $R$ term,
has all the characteristics of a proper impact factor: it is composed of terms that
can be unambiguously assigned to each external leg of the amplitude, and it is
completely consistent with the interpretation of the impact factor as the action of
two `jet operators', as defined in \eq{jetfactors}, on the hard part of the amplitude
$\widehat{H}_{aa}$. The second line of \eq{c2imp}, on the other hand, clearly does
not admit an interpretation as an `impact factor', which should be associated with
pure color-octet exchange, and should depend only on the identity of the particles being
scattered on either side of the $t$-channel Reggeized propagator. On the contrary,
the second line of \eq{c2imp} contains the color operator ${\bf T}^2_s$, which mixes
the representations being exchanged in the $t$ channel, and depends on the identity
of all the four particles participating in the scattering. Furthermore, although real, the 
second line in \eq{c2imp} originates from the phase factor in ${\cal Z}_{\bf 1}$, which
is difficult to reconcile with the reality properties required by high-energy factorization.

Armed with these considerations, we propose to define the impact factors precisely 
as the set of terms in \eq{c2imp} that arise from the action of the `jet operators' in 
\eq{jetfactors} on the hard coefficients. At two loops this gives
\beq
  \widetilde{C}_a^{(2)} \, = \, \frac{1}{2} Z^{(2)}_{1, {\bf R}, aa} - \frac{1}{8} 
  \left( Z^{(1)}_{1, {\bf R}, aa} \right)^2 + \frac{1}{4} Z^{(1)}_{1, {\bf R}, aa} \, 
  \RE \left[ \widehat{H}^{(1), 0, [8]}_{aa} \right] \, + {\cal O} \left( \epsilon^0 \right) \, .
\label{newC}
\eeq
Correspondingly, we propose to define the non-factorizing remainder $R$ at two 
loops as
\beqa
  \widetilde{R}^{(2),0 , [8]}_{ab} & = & - \, \frac{\pi^2 (K^{(1)})^2}{H^{(0),[8]}_{ab}}
  \Bigg[ \left( ({\bf T}^2_{s, ab})^2 H^{(0)}_{ab} \right)^{[8]} - {\cal C}_{{\rm tot},\,ab}
  \left({\bf T}^2_{s, ab}H^{(0)}_{ab} \right)^{[8]} \nonumber \\
  && \hspace{5mm} - \left(\frac{1 + \kappa_{ab}}{2}
  N_c^2 - \frac{{\cal C}^2_{{\rm tot},\,ab}}{4} \right) H^{(0),[8]}_{ab} \Bigg] 
  \, + {\cal O} \left( \epsilon^0 \right) \, .
\label{rest}
\eeqa
We note that  \eq{rest} has no single pole terms, which is a consequence of the 
fact that it arises ultimately from the square of the phase factor in \eq{facZ1}. The
expression in \eq{rest} is still somewhat formal, but it can easily be made explicit,
for each parton species, upon picking specific color bases for the various 
amplitudes. Working in the orthonormal bases described in detail in 
Ref.~\cite{Beneke:2009rj}, we get 
\beq
  \widetilde{R}^{(2), 0, [8]}_{qq} \, = \, \frac{\pi^2}{4 \epsilon^2}
  \left(1 - \frac{3}{N_c^2} \right) \, , \quad \, 
  \widetilde{R}^{(2), 0, [8]}_{gg} \, = \,  - \, \frac{ 3 \pi^2}{2 \epsilon^2} \, , \quad \,
  \widetilde{R}^{(2), 0, [8]}_{qg} \, = \, - \, \frac{\pi^2}{4 \epsilon^2} 
  \,. \nonumber
\eeq
In particular, one can verify, using the results of Ref.~\cite{Catani:1998bh,Sterman:2002qn}, 
that $\widetilde{R}^{(2),0,[8]}_{qg}$, together with the impact factors as defined in \eq{newC}, 
accounts for all the poles of the two-loop quark-gluon scattering amplitude.

Note that, had we used the hypothesis of Regge factorization without a non-factorizing 
remainder, as was done in Ref.~\cite{DelDuca:2001gu}, we would have found a 
mismatch between the quark-gluon scattering amplitude and the one predicted
by the Regge factorization formula, \eq{ReggeFact}, without the remainder $R$.
That mismatch may be quantified by the function,
\beqa
  \Delta_{(2),0,[8]} & = & \frac{M^{(2),0}_{qg}}{H^{(0),[8]}_{qg}} - 
  \bigg[C^{(2)}_q + C^{(2)}_g + C^{(1)}_q C^{(1)}_g - \frac{\pi^2}{4} 
  \left(1 + \kappa \right) (\alpha^{(1)})^2 \bigg] \nonumber \\
  & = & \frac{1}{2}\bigg[\widetilde{R}^{(2), 0, [8]}_{qg} - \frac{1}{2} 
  \left(\widetilde{R}^{(2), 0, [8]}_{qq} + 
  \widetilde{R}^{(2), 0, [8]}_{gg} \right)\bigg] \, .
\label{delta}
\eeqa
Using data from our chosen color basis~\cite{Beneke:2009rj}, we may evaluate
explicitly \eq{delta}, finding
\beq
  \Delta_{(2),0,[8]} \, = \, \frac{\pi^2 (K^{(1)})^2}{2} \bigg[\frac{3}{2} \left(\frac{N_c^2 + 1}{N_c^2}
  \right) \bigg] \, = \, \frac{\pi^2}{\eps^2} \frac{3}{16} \left(\frac{N_c^2 + 1}{N_c^2} \right) \, .
\label{findelta}
\eeq
\eq{findelta} is in complete agreement with the discrepancy found in 
Ref.~\cite{DelDuca:2001gu}, and explains the origin of the problem, as arising 
from the mixing of color representations and the phase factors that are required
by soft-collinear factorization.

Proceeding to three-loop order, one would expect that matching the single-logarithmic terms 
of Eqs.~(\ref{ReggeFact}) and (\ref{AmpExpansion}) should allow us to obtain a universal 
expression for the three-loop Regge trajectory $\alpha(t)$. As predicted in~\cite{DelDuca:2011ae,
Bret:2011xm}, however, a direct comparison yields a non-universal result, in agreement with \eq{NNLL}. To illustrate the situation, we quote here the triple pole contribution, which is where 
the leading factorization-breaking effects arise, and which is completely determined by 
soft factors. A detailed discussion of the complete three-loop predictions for impact factors and
for the Regge trajectory is left to Ref.~\cite{us}. The non-universal result for the three-loop 
Regge trajectory reads, at this level,
\beqa
  \alpha^{(3)} & = & C_A K^{(3)} + \frac{\pi^2 (K^{(1)})^3}{2} 
  \bigg[{\cal C}_{{\rm tot}, ab} N_c \left( {\bf T}^2_{s, ab} \right)_{[8],[8]}
  - \frac{{\cal C}^2_{{\rm tot}, ab} N_c}{4} + \frac{1 + \kappa_{ab}}{2} N_c^3 \nonumber  \\
  && \, - \, \frac{1}{3} \sum_n \left(2 N_c + {\cal C}_{[n]} \right) \left| 
  \left({\bf T}^2_{s, ab} \right)_{[8], n} \right|^2 \bigg] -\frac{1}{2} R^{(3), 1, [8]}_{ab}
  + {\cal O} \left( \epsilon^{-2} \right) \, ,
\label{alpha3}
\eeqa
where the sum on the second line runs over all color representations that can be exchanged 
in the $t$ channel. Once again, we recognize that the first term has the appropriate universality properties, and indeed corresponds to the all-order ansatz for infrared-singular contributions 
to $\alpha(t)$ first given in~\cite{Korchemskaya:1996je} and then reproduced
in~\cite{DelDuca:2011ae,Bret:2011xm}. The other terms in \eq{alpha3} are clearly of a 
non-universal nature, and it is appropriate to attribute them to the non-factorizing 
remainder $R$. We define then
\beqa
 \widetilde{\alpha}^{(3)} & = & K^{(3)} N_c + {\cal O} \left( \epsilon^0 \right) \, , 
 \nonumber \\ 
 \widetilde{R}^{(3), 1, [8]}_{ij} & = & \pi^2 (K^{(1)})^3 \, \bigg[{\cal C}_{{\rm tot}, ij}
 N_c \left({\bf T}^2_{s, ij} \right)_{[8],[8]} - \frac{{\cal C}^2_{{\rm tot}, ij} N_c}{4}
 \label{3loopR} \\ && \hspace{1cm} 
 + \, \frac{1 + \kappa}{2} N_c^3 - \frac{1}{3} \sum_n \left( 2 N_c + {\cal C}_{[n]} 
 \right) \left| \left( {\bf T}^2_{s, ij} \right)_{[8],n} \right|^2 \bigg] 
 + {\cal O} \left( \epsilon^{-2} \right) \, . 
 \nonumber
\eeqa
We emphasize that \eq{3loopR} is an absolute prediction for single-logarithmic terms
of high-energy three-loop quark and gluon amplitudes, which is of purely infrared
origin and does not rely upon any input from lower-order finite contributions to the 
amplitudes. Similar results can be derived for double and single poles of $R^{(3),1,[8]}$,
and will be described in~\cite{us}, but they require progressively more detailed information
from finite-order calculations. If we introduce the appropriate color factors in \eq{3loopR}, 
working as before in the color bases if ~\cite{Beneke:2009rj}, we obtain the explicit results
\beqa
  \widetilde{R}^{(3), 1, [8]}_{qq} & = & \left(\frac{\as}{\pi} \right)^3 \frac{\pi^2}{\epsilon^3} \, 
  \frac{2 N_c^2 - 5}{12 N_c} \, , \nonumber \\
  \widetilde{R}^{(3), 1, [8]}_{gg} & = & - \, \left(\frac{\as}{\pi} \right)^3 \frac{\pi^2}{\epsilon^3} \,
  \frac{2}{3} \, N_c \, , \\
  \widetilde{R}^{(3), 1, [8]}_{qg} & = & - \, \left(\frac{\as}{\pi} \right)^3 \frac{\pi^2}{\epsilon^3} \,
  \frac{N_c}{24} \, , \nonumber
\label{explR3}
\eeqa
which can be consistently used in \eq{ReggeFact}, provided one also substitutes our new
definitions of the impact factors and of the Regge trajectory, as given in Eqs.~(\ref{newC}) 
and (\ref{3loopR}).

\section{Perspective}
\label{discu}

High-energy factorization and soft-collinear factorization are often studied with 
different techniques, and applied to different kinematical domains. One might however 
argue that, in some sense, high-energy logarithms are a special class of infrared logarithms, 
arising when certain scales of the problem become much smaller than other ones. The wealth 
of techniques which are routinely applied to study the soft approximation becomes then
available to study the high-energy limit as well. This viewpoint is of course well known: it 
was pioneered in~\cite{Korchemsky:1993hr,Korchemskaya:1994qp,Korchemskaya:1996je}, 
further developed in~\cite{DelDuca:2011ae,Bret:2011xm} using more recent technical 
developments\footnote{See also~\cite{Balitsky:1998ya,Balitsky:2001gj,Kucs:2003ei}, for 
other analyses in a similar spirit.}, and indeed it is a crucial ingredient of the methods 
recently proposed in~\cite{Caron-Huot:2013fea}. 

In this paper, we have presented some preliminary results that follow from a detailed 
comparison of the two factorizations, order by order in perturbation theory. We have 
considered specifically quark and gluon amplitudes in QCD, though we emphasize 
that very similar results could easily be derived for other gauge theories (for example 
for the interesting case of $N = 4$ Super-Yang-Mills theory, where our results would 
concern contributions beyond the planar limit). Building upon the detailed factorization 
derived in~\cite{DelDuca:2011ae,Bret:2011xm}, we have used soft-collinear 
techniques  in the high-energy limit to explore the limitations of the Reggeization picture, 
as realized under the assumption that only isolated poles arise in the complex angular 
momentum plane. As discussed most recently in~\cite{Caron-Huot:2013fea}, it is 
understood that this picture must break down, as Regge cuts arise at sufficiently high 
orders in perturbation theory. Soft-collinear factorization provides a powerful tool to 
explore the onset of these new effects, as it gives explicit expressions for the (infrared 
singular) contributions to the amplitudes that break the simplest form of Reggeization, 
starting at two loops for non-logarithmic terms, and continuing to higher orders at NNLL 
accuracy. 

Comparing the two factorizations, we have noted that infrared constraints provide explicit 
expressions for the impact factors and for the Regge trajectory, which receive clearly 
non-universal contributions starting at two loops for the impact factors and at three loops 
for the Regge trajectory. We have proposed to collect the universal terms by properly 
redefining the impact factors and the Regge trajectory order by order, and to gather 
the non-universal contributions into a non-factorizing remainder function. Using our 
definitions, we have been able to reconstruct the origin of the discrepancy from 
high-energy factorization discovered in~\cite{DelDuca:2001gu}, which arises in our 
framework as a linear combination of the non-factorizing remainders of two-loop 
quark-quark, gluon-gluon and quark-gluon amplitudes. Furthermore, at the three-loop 
level, we have given a precise definition of the Reggeization-breaking factors which 
provide non-universal single-logarithmic contributions to the amplitudes, and we have 
explicitly computed these terms for all relevant QCD amplitudes. 

We emphasize that, while in this letter we have given explicitly only the leading singular 
contributions to the non-factorizing remainders, similar expressions can be derived also 
for subleading terms, and they will be presented in detail in Ref.~\cite{us}. Similarly, we 
note that in the present paper we have focused on the real parts of the amplitudes,
and our results mostly take the form of constraints on high-energy factorization arising
from soft-collinear universality. In~\cite{us}, we will also consider imaginary parts of
amplitudes, and we will show that high-energy factorization, in turn, provides important 
constraints on the soft, collinear and hard functions entering the soft-collinear factorization 
formula. Finally, we note that we have concentrated here on four-point amplitudes, for the
sake of simplicity. The results of Refs.~\cite{DelDuca:2011ae,Bret:2011xm}, however, apply 
also to multi-parton amplitudes in multi-Regge kinematics, where a high-energy factorized
expression for the amplitude is also available. In view of the phenomenological relevance
of this kinematical situation to LHC physics searches~\cite{Andersen:2009nu}, it will be 
interesting to apply our techniques to explore the boundaries of high-energy factorization
in this regime as well.

\vspace{1cm}

{\large{\bf Acknowledgements}}

\vspace{2mm}

\noindent We thank S. Caron-Huot for useful discussions. This work was supported 
by MIUR (Italy), under contracts 2006020509$\_$004 and  2010YJ2NYW$\_$006; 
by the Research Executive Agency (REA) of the European Union, through the Initial 
Training Network LHCPhenoNet under contract PITN-GA-2010-264564, and by the 
ERC grant 291377 ``LHCtheory: Theoretical predictions and analyses of LHC physics: 
advancing the precision frontier''. VDD thanks CERN and the Institut f\"ur Theoretische 
Physik, Universit\"at Z\"urich, and LM thanks CERN, NIKHEF, and the Higgs Center for 
Theoretical Physics at the University of Edinburgh for hospitality and support during the 
completion of this work.

\end{document}